\documentclass[a4paper,11pt]{article}
\usepackage{pos}
\usepackage{subcaption}
\usepackage{cleveref} 
\usepackage{graphicx}
\usepackage{sidecap}
\usepackage{float}
\usepackage{makecell}
 \usepackage{wrapfig} 
\usepackage{multirow}
\usepackage{booktabs}
\usepackage{dcolumn}
\newcolumntype{d}[1]{D{.}{.}{#1}}

\def\orcid#1{\kern .08em\href{https://orcid.org/#1}{\includegraphics[keepaspectratio,width=0.7em]{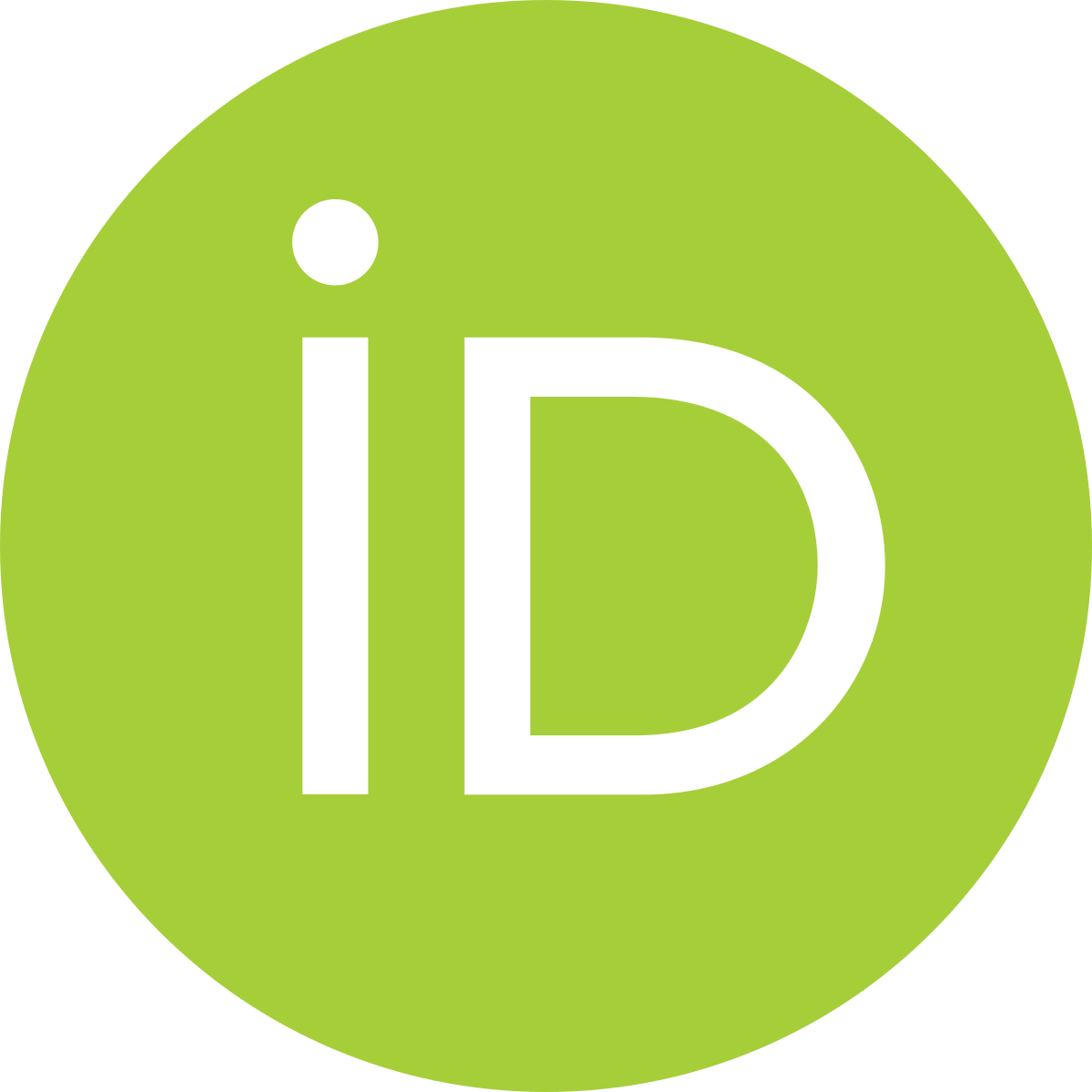}}}

\title{\textbf{Effects of closely spaced thresholds on line shapes with near-threshold enhancement}}

\ShortTitle{Effects of closely spaced thresholds on line shapes...}

\author*{Exan John D.F. Carpio\orcid{0009-0001-6398-2959}}
\author{Denny Lane B. Sombillo\orcid{0000-0001-9357-7236}}

\renewcommand{\printHeadAuthors}{EJDF Carpio\orcid{0009-0001-6398-2959} and DLB Sombillo\orcid{0000-0001-9357-7236}}

\affiliation{National Institute of Physics, University of the Philippines Diliman,\\
Quezon City 1101, Philippines}

\emailAdd{edcarpio1@up.edu.ph}
\emailAdd{dbsombillo@up.edu.ph}

\abstract{ Hidden-charm pentaquarks were first experimentally detected by LHCb in 2019, one of which is the exotic $P_{c\bar{c}}(4312)^{+}$ state. The nature of this state remains uncertain which may be attributed to the proximity of this observed enhancement to the meson-baryon thresholds. In this study, we introduce a coupled-channel approach using three-channel separable potential model to account for near-threshold effects on the observed signal. In particular, we assign $\Sigma_c^+\bar{D}^0$ and $\Sigma_c^{++}D^-$ as higher-mass channels for $P_{c \bar{c}}(4312)^+$ state. Moreover, this allows us to propose four pole configurations, where either a bound state or virtual state pole were placed near the higher-mass thresholds. Using this scheme, we study the near-threshold effects on the transition of different pole configurations across the unphysical Riemann sheets and examine their effects on the amplitude line shape. We found out that $P_{c \bar{c}}(4312)^+$ state may be interpreted as a virtual state below the $\Sigma_c^+\bar{D}^0$ which is consistent with our initial modeling assumptions. Our result also conforms with the previous analysis done by Joint Physics Analysis Center (JPAC) \cite{Fernandez-Ramirez:2019koa}, which leans toward similar interpretation. This may indicate that a model-dependent framework employing a relatively simple model, such as a separable potential, can somehow complement with some of the widely used parameterizations.

}

\FullConference{
The 21st International Conference on Hadron Spectroscopy and Structure (HADRON2025)\\
27-31 March, 2025\\
Osaka University, Japan
}

\begin{document}
\maketitle

\section{Introduction}
Identifying the nature of exotic $P_{c\bar{c}}(4312)^+$ state remains unsettled due to the near-threshold effects which may caused this observed enhancement \cite{LHCb:2019kea}. This led to several interpretation such as, hadronic molecule \cite{Guo:2019kdc,Du:2021fmf}, compact pentaquark \cite{Cheng:2019obk}, pole-based enhancement \cite{Santos:2024bqr} or purely kinematical effects. These competing interpretations make it a challenging task to establish a universal consensus within the community.

In this study, we investigate the influence of nearby threshold effects on the line shapes of exotic pentaquark states. To model these phenomena, we employ a coupled-channel formalism with a separable potential approach to effectively captures the dynamics associated with threshold interactions. Within this framework, three distinct thresholds with unequal masses are relevant to the $J/\psi p$ invariant mass spectrum.  This setup enables a detailed analysis of the pole trajectories in the complex energy plane as the coupling parameters are varied.

\section{Coupled-channel formalism}
To systematically study how the nearby thresholds influence the observed enhancements, we adopt a coupled-channel formalism. This approach allows us to track how different pole configurations emerge under varying interaction strengths. The $T-$matrix, central to scattering theory, encapsulates how incoming particles interact and transform due to their mutual potentials, which takes the form of
\begin{align}\label{1}
    t_{a b}\left(p, p';E\right)= v_{a b}\left(p, p'\right) + \sum_{\gamma}^{} \int_{0}^{\infty} d  p'' p''^2  v_{a \gamma}\left(p, p''\right)  \frac{1}{E -\epsilon_\gamma-\frac{p''^2}{2\mu_\gamma}}   t_{\gamma b}\left(p'', p'; E\right),
\end{align}
where indices $a$ and $b$ represents the three channels. The mutual potential, $ v_{a b}\left(p, p'\right)$ takes the separable form, $  v_{a b}\left(p, p'\right)= \lambda_{a b} f_a \left(p\right) f_b\left(p'\right)$, with $f_a \left(p\right)$ represents the form factor.  In our case, it helps us model the three relevant channels, $J/\psi p$, $\Sigma^+_c \bar{D}^0$, and $\Sigma^{++}_c {D}^-$ interfere near threshold energies By applying separable potentials with form factors, we derive analytic expressions for the scattering amplitude
\begin{align}\label{4}
T_{a b}= -\pi \sqrt{\mu_a \mu_b  k_a k_b} f_a\left(p\right) \tau_{ab} \left(E\right) f_b \left(p'\right)\Theta\left(E-\epsilon_a\right) \Theta\left(E-\epsilon_b\right),
\end{align}
where $\left[\tau^{-1}\left(E\right)\right]_{ab}=\left(\lambda^{-1}\right)_{ab}-\delta_{ab} I_a \left(E\right)$. This framework allows us to explore how pole trajectories respond to varying couplings, which is a core aspect of the results presented in Section~\ref{sec3}.
\section{Results and discussion}\label{sec3}
\begin{figure}[ht]
  \centering
    \centering
    \subfloat[]{\includegraphics[width=\textwidth]{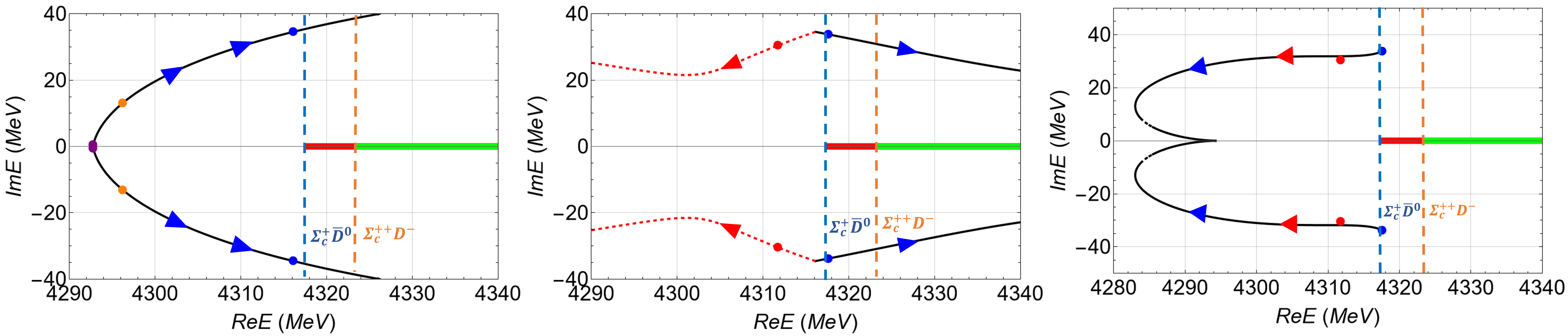}\label{ET_M1}}\\
    \subfloat[]{\includegraphics[width=\textwidth]{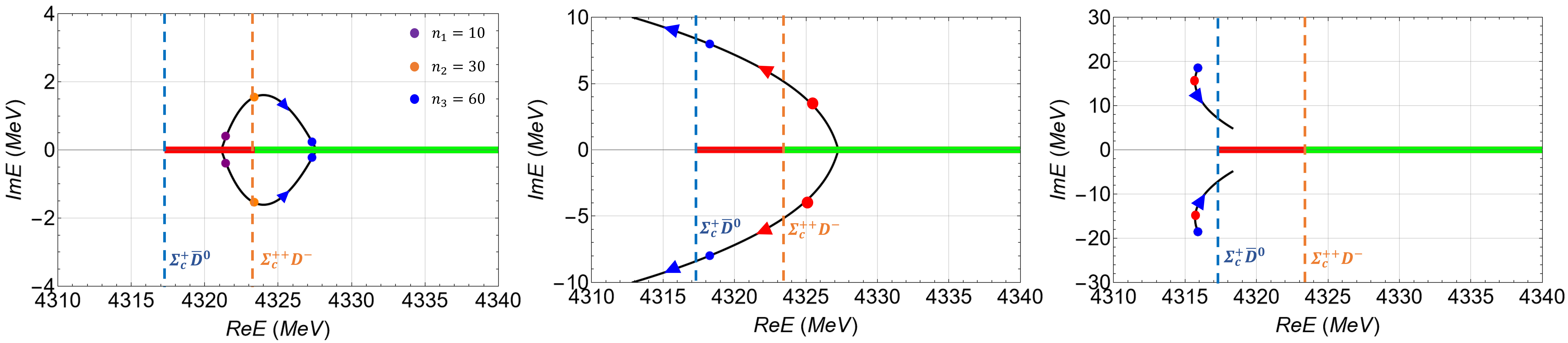}\label{ET_M2}}
    \caption{Energy pole trajectory as the coupling strength parameters are systematically increased. From left to right panels, we schematically activated the coupling parameters as follows[$\left(\lambda_{12}>0\right)$, $\left(\lambda_{12},\lambda_{23} >0\right)$, and $\left(\lambda_{12},\lambda_{23}, \lambda_{13} >0\right)$]. Each successive panel incorporating an additional coupling parameter. (a) Model 1: Virtual state pole in second channel. (b) Model 3: Virtual state pole in the third channel.}
    \label{ET}
\end{figure}

In this work, four models were used to examine the near-threshold effects on pole trajectories and amplitude line shapes. Models 1 and 2 introduce a virtual state and a bound state pole, respectively, below the second thresholds, whereas, Models 3 and 4 follow an analogous pole configurations between the second and third thresholds. A virtual pole in channel 1 below the $J/\psi p$ threshold is placed by default, serving as common feature to all the models. To study the internal structure of a given resonance state, a pole analysis was imposed. Under this framework, we investigated the poles of the scattering amplitude by studying their trajectory in the complex energy plane as the coupling strengths are varied, as shown in Fig.~\ref{ET}. In the first panel of Fig.~\ref{ET_M1}, the virtual state pole starts its trajectory above the second threshold, and threshold, and as the $\lambda_{12}$ coupling increases, a dynamically generated

\begin{figure}[H]
    \centering
    \includegraphics[width=0.7\linewidth]{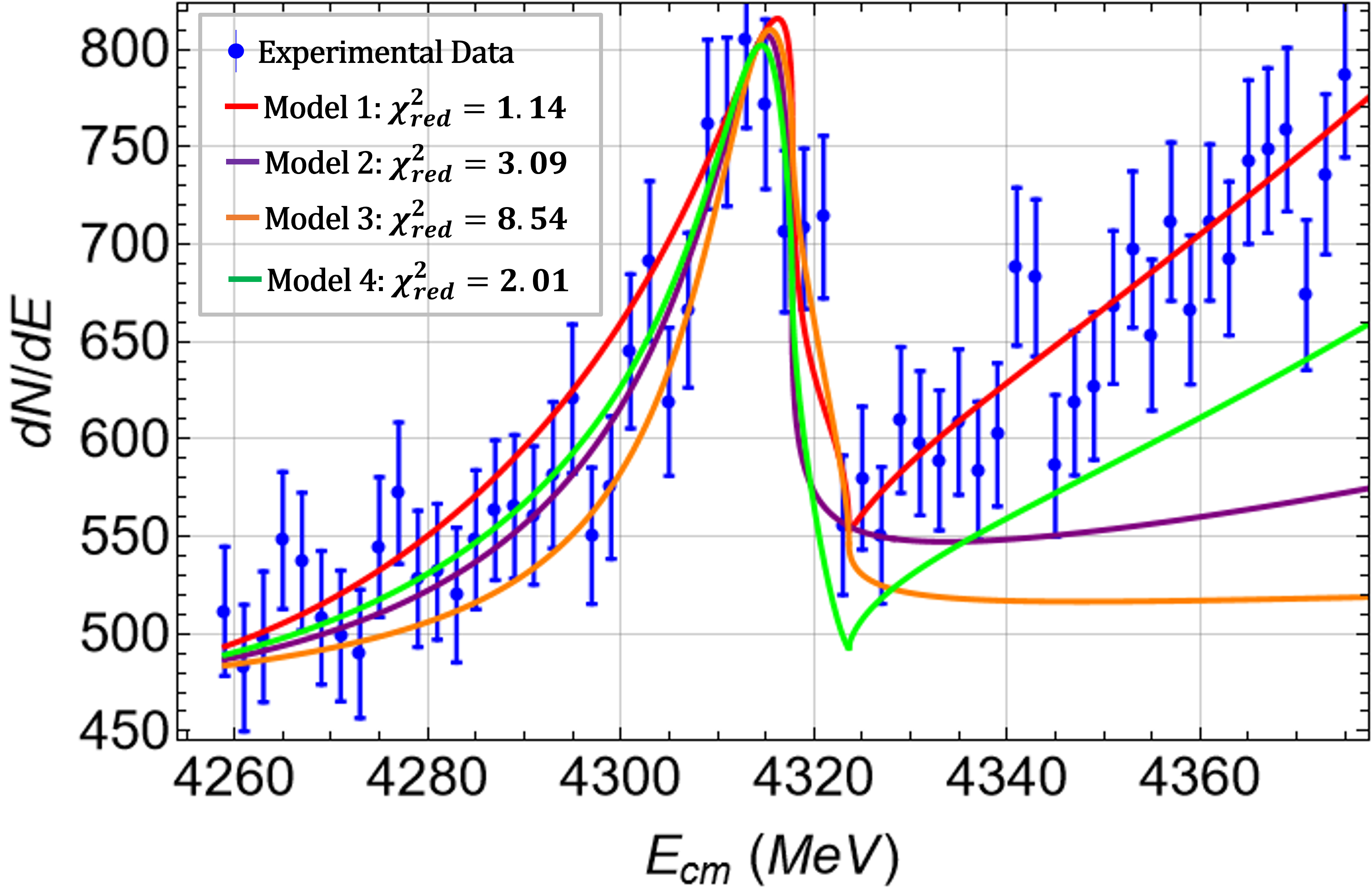} 
     \caption{Comparison of the event mass distributions from different pole configurations with the $J/\psi p$ invariant mass distribution of the LHCb for exotic $P_{c\bar{c}}(4312)^+$state}
    \label{experimental}
\end{figure}

\noindent
 resonance emerges, which continues beyond the second threshold. These resonance poles (indicated by blue arrows) continue their trajectories above the third threshold as the $\lambda_{23}$ coupling varies, in contrast to the shadow poles (red arrows), which traverse below the first and second thresholds. However, as we activate all the couplings, the resonance and shadow poles seem to move below the real axis of the second threshold. On the other hand, the leftmost panel of Fig.~\ref{ET_M1} exhibits a pair of resonance poles which transition to a different Riemann sheet by crossing the real axis above the third threshold. These poles, together with their corresponding shadow poles, emerge above the third threshold and trace their trajectories below the third threshold, as shown in the second and third panels of Fig.~\ref{ET_M2}. These trends exhibited by the two models support our hypothesis about threshold proximity influencing the pole behavior. Moreover, the effects of the shadow poles can be seen in Fig.~\ref{experimental}, where a dip near the third threshold for Model 1 is observed. This model captures of line shape of experimental data with $\chi^2_{red}=1.12$.
 In hindsight, model 3 exhibits a plateau- like trend above the third threshold, and such behavior can be attributed to the closeness of shadow poles from the physical region, leading to a $\chi^2_{red}$ of 8.54. It is worth noting that models 2 and 4, with bound state poles below the second and third thresholds, respectively, mimic the $\sim$4312 MeV peak near the $\Sigma^+_c\bar{D}^0$ threshold, but fail to match the behavior beyond the third threshold. Hence, one can conclude that model 1 accurately captures the sharp enhancement at $\sim$4312 MeV and the line shape trend above the third threshold, aligning with the virtual state interpretation. Interestingly, this result is in agreement with the previous findings of Ref.~\cite{Fernandez-Ramirez:2019koa}, despite the completely different methodological formalism implemented.

\section{Conclusion and outlook}
In this work, we probed the possible interpretation of the $P_{c\bar{c}}(4312)^+$ by using a model-dependent framework. Four pole configurations were implemented where either a bound state or a virtual pole is placed near the thresholds. Using such scheme, we were able to show the effects of closely spaced thresholds on the pole behavior and scattering amplitude. Furthermore, the results suggest that the model 1 best reproduces the peak at $\sim$ 4312 MeV and the overall line shape trend. Hence, the exotic $P_{c\bar{c}}(4312)^+$ enhancement may be interpreted as a virtual state pole below the $\Sigma^+_c \bar{D}^0$ threshold. Looking ahead, incorporating model-independent framework together with a more sophisticated approach represents a promising direction for future analyses.

\acknowledgments
EJDFC acknowledges the scholarship support provided by the DOST-ASTHRDP.

\noindent



\bibliographystyle{JHEP}
\bibliography{bib_Carpio}

\end{document}